# Magnetic field control of ferroelectric polarization and magnetization of LiCu$_2$O$_2$ compound[*]

Qi Yan (齐岩) and Du An (杜安)[†]

*College of Sciences, Northeastern University, Shenyang 110819, China*

**Abstract**

A spin model of LiCu$_2$O$_2$ compound with ground state of ellipsoidal helical structure has been adopted. Taking into account the interchain coupling and exchange anisotropy, we focus on the magnetoelectric properties in a rotating magnetic field and perform the Monte Carlo simulation on a two-dimensional lattice. A prominent anisotropic response is observed in the magnetization and polarization curves, qualitatively coinciding with the behaviors that detected in the experiment. In addition, the influences of the magnetic field with various magnitudes are also explored and analyzed in detail. As the magnetic field increases, a much smoother polarization of angle dependence is exhibited, indicating the strong correlation between the magnetic and ferroelectric orders.

[*]Project supported by Academic Scholarship for Doctoral Candidates of China (Grant No. 10145201103), the Fundamental Research Funds for the Central Universities of China (Grant No. N110605002) and Shenyang Applied Basic Research Foundation (Grant No. F12-277-1-78) of China.

[†]Corresponding author. E-mail: duanneu@163.com

# 1. Introduction

Multiferroics, which possess magnetic and ferroelectric orderings in the same material, yield abundant physical mechanisms and broad application prospects. In recent years, the magnetism-driven ferroelectricity discovered in frustrated multiferroic materials has attracted considerable attention.[1-3] In these compounds, spiral spin states are usually found and give rise to surprising physical properties, such as flop, reversal and rotation of the electric polarization in an external magnetic field.[4-7] This brings a hopeful prospect of controlling charges by applied magnetic fields and using this to construct new forms of multifunctional devices.

The $LiCu_2O_2$ multiferroic compound is a prototype of one-dimensional $S=1/2$ spiral magnets.[8-15] It has an orthorhombic crystal structure with equal number of magnetic $Cu^{2+}$ and nonmagnetic $Cu^{1+}$ ions. The magnetic $Cu^{2+}$ ions are located at the center of edge-sharing $CuO_4$ plaquettes and form a zigzag like spin-chain structure along the $b$ axis. The Cu-O-Cu angle is nearly 94°.[16-19] As a result, according to the Kanamori-Goodenough rule, the nearest-neighbor ferromagnetic (FM) coupling $J_2$ and next-nearest-neighbor antiferromagnetic (AFM) coupling $J_4$ are expected along the chain direction $b$ axis. Such competing interactions will lead to frustration and an incommensurate helimagnetic state in this quantum spin system. Spontaneous electric polarization has been detected when the helical order sets in, and shows sensitive response to the applied field.

In order to reveal the origin of the polarization, many experiments have been carried out to study the magnetic structures of $LiCu_2O_2$. A former neutron scattering study has indicated the helical magnetic order in the $ab$ plane.[16] However, it contradicts with the spin-current model or the inverse Dzyaloshinskii-Moriya interaction since spontaneous electric polarization emerges along the $c$-axis. Later, Park et al. based on the polarized neutron scattering experiments proposed the $bc$-spiral plane,[11] which is partially supported by Seki et al., but the quantitative calculation of the intensity of polarized neutron reflections shows a prominent discrepancy.[20] Recently, by combining studies of neutron diffraction and Li-NMR results, Sato et al. proposed an ellipsoidal helical spin structure with the helical axis tilted by about 45° from the *a*- or *b*- axis within the *ab* plane.[21,22] This new noncollinear spin configuration receives strong evidence support from the very recent experiments performed by Li et al.[23] They implement the measurements of the dielectric constant and spontaneous polarization in nearly untwinned LiCu$_2$O$_2$ single crystals. The highly anisotropic responses observed in different magnetic fields are perfectly interpreted by using this 45°-tilt spin model.

Although a vast amount of experiments upon LiCu$_2$O$_2$ have been done,[24-28] theoretical studies on its thermodynamic properties or field dependent behaviors are still sparse and unsatisfied owing to inadequate information on ground states. The ellipsoidal helical spin structure suggested by Sato et al. provides us the new clues to unveil the underlying mechanism in this multiferroic compound. In this paper, concentrating on magnetoelectric coupling effect, the magnetization and polarization under a rotation of the magnetic field are investigated by employing Monte Carlo method. The results demonstrate the strong coupling between the ferroelectric and magnetic orders in this spiral magnet, and in some respects are qualitatively in good agreement with that of experiment.

## 2. Model and Simulation

In LiCu$_2$O$_2$, there are two linear Cu$^{2+}$ chains, which propagate along the *b* axis and form a zigzag ladderlike structure [Fig. 1]. The Hamiltonian of the system is written as

$$H = \sum_{i,j}\left[ J_1(\mathbf{S}_i \cdot \mathbf{S}_j)_{XXZ} + J_2(\mathbf{S}_i \cdot \mathbf{S}_j)_{XXZ} + J_4(\mathbf{S}_i \cdot \mathbf{S}_j)_{XXZ} + J_\perp(\mathbf{S}_i \cdot \mathbf{S}_j)_{XXZ} \right] - \sum_i D_i(\mathbf{S}_i \cdot \mathbf{e}_i)^2 - \mathbf{h} \cdot \mathbf{S}_i - \mathbf{h}_e \cdot \mathbf{P}_i \quad , \quad (1)$$

where $(\mathbf{S}_i \cdot \mathbf{S}_j)_{XXZ} = S_i^x S_j^x + S_i^y S_j^y + \Delta S_i^z S_j^z$ with unit vectors of the classical spin components $(S_i^x, S_i^y, S_i^z)$. $\Delta \leq 1$ is an exchange anisotropy and theoretically expected to stabilize a vector chiral order.[26] Actually, in the edge-sharing spin-chain compounds substantial exchange anisotropies have been detected.[29] In order to yield results that much close to those of experiments, the value of $\Delta = 0.7$ is adopted.

Considering the 45°-tilt spin model proposed by Sato et al. (Fig. 2), a planar anisotropy in the [110] direction has been added in the Hamiltonian. $D_i$ represents the magnitude of the magnetic anisotropy. $e_i$ is a unit vector and represents the direction of magnetic anisotropy. For the sake of effectively pinning the spiral axis at the diagonal of $ab$ plane, a large negative $D_i = -5$ is made. In addition, this planar anisotropy will lead to a strong spin coupling along $c$ axis, which is found to be essential for inducing electric polarization.[30] Here $h$ is the external magnetic field including an extra factor $g\mu_B$ and $h_e$ is the external electric field applied along the $c$ axis. The average magnetization is evaluated as

$$m = \frac{M}{N} = \frac{1}{N}\sum_i S_i, \qquad (2)$$

with $N$ the total number of spin sites and $M$ the total magnetization.

According to the spin current model, or equivalently, the inverse Dzyaloshinskii-Moriya interaction, the polarization $P_i$ induced by two neighboring spins ($S_i$ and $S_j$) is formulated as

$$P_i = A e_{ij} \times (S_i \times S_j) \qquad (3)$$

where $e_{ij}$ denotes the vector connecting the two sites of $S_i$ and $S_j$, namely in the direction of magnetic modulation vector along the $b$ axis. $A$ is a proportional constant determined by the spin-exchange and the spin-orbit interactions as well as the possible spin-lattice coupling term. It is assumed to be unity here. Thereby the third item in Hamiltonian Eq. (1) corresponds to electric energy.

The low-dimensional $LiCu_2O_2$ compound contains multifold spin interactions as shown in Fig. 1. It is well accepted that $J_2$ is ferromagnetic and $J_4$ is antiferromagnetic. The estimated ratio $|J_4/J_2|$ approximately varies from 0.5 to 0.65, which has been experimentally verified.[16,28] However, the relative magnitudes of interchain interactions $J_1$ and $J_\perp$ remain controversial. Based on inelastic neutron scattering experiments and spin wave theory, Masuda et al.[17,18] extract the relevant exchange constants presented in this spin system. Their data suggest a strong AF "rung" interaction $J_1$ and a weak interchain coupling $J_\perp$, in contrary to the conclusion obtained by Drechsler et al.[13] In this paper, we adopt the suggestion of Masuda et al. here. Our previous study showed that this enhanced antiferromagnetic coupling $J_1$ would cause a drastic competition with the antiferromagnetic bond $J_4$ and lead to the modification of the modulated period of the incommensurate spiral order, and more important, the thermodynamic behaviors under this condition were in qualitatively good agreement with that of experiment.[31] Therefore, according to the exchange parameters from neutron scattering and our previous investigation on this model, the parameters are lined out as follow: $J_1$=3.4, $J_2$= -6.0, $J_4$=3.0 and $J_\perp$=0.9. One can note that the ratio $|J_4/J_2|$=0.5 here is also in the range provided by the experiment.

Since $LiCu_2O_2$ compound possesses a weak $J_\perp$ interaction and the spiral order is formed along the chain, the number of the zigzag chains is fixed at $L'$=10. For the length of a zigzag chain, we have carefully examined the size effects and find that $L$=100 is large enough to eliminate it. Hence the lattice size in the present simulation is 100×10. It is assumed that $a$, $b$ and $c$ axes are respectively, along the directions of [100], [010], and [001]. The standard Monte Carlo simulation is employed and performed on this $L \times L'$ lattice with periodic boundary conditions. The spin is updated according to the Metropolis algorithm. For every T, the initial 50000 Monte Carlo steps (MCS) are discarded for equilibration, and then the results are obtained by averaging 15000 data. Each data is collected at every 10 MCS.

The final results are obtained by averaging twenty independent data sets obtained by selecting different seeds for random number generation.

## 3. Results and Discussions

Similar to the measurement process in experiment, the system is initially polarized by an electric field $h_e$=0.5 along the $c$ axis, which is implemented under the condition of the zero magnetic field cooling (ZFC). When the system is polarized down to a very low temperature $T$=0.01, a single magnetoelectric domain will be produced with magnetic modulation vector along the chain direction. Based on the cubic symmetry of magnetic structure, the spiral axes are along [110], [$\bar{1}\bar{1}$0], [1$\bar{1}$0], or [$\bar{1}$10] with equal probability. The choice depends on the sets of parameters. In the present simulation, we have considered a large negative magnetic anisotropy $D_i = -5$ in the diagonal direction of *ab* plane, therefore the ground state consisting of the incommensurate spiral structure with its spiral axis along the [110] direction can be expected. As shown in Fig. 2(a), if all the sites in the chains are moved to one point, the spin vectors would almost lie on the vice diagonal plane of the cube and form an ellipsoidal shape, which well reproduce the ground state configuration described in Ref. [21,22]. In Fig. 2(b), the snapshot of the spin configuration in a single chain also demonstrates a vivid picture of the low-temperature spiral magnetic order.

In order to analyze the spiral spin state accurately, the spin structure factor is evaluated along the spin chain via a Fourier transformation of the spin correlation function written as[32]

$$S(q) = \sum_{i,r} \cos(q \cdot r) \langle S_i \cdot S_{i+r} \rangle, \qquad (4)$$

where *q* is wave vector. *r* is calculated in units of distance between two nearest-neighbor correlated spins. The spin structure factor, which can be measured by a diffuse neutron-scattering experiment, provides the direct information about the spin ordering at finite temperature through the position of its maximum. In Fig. 2(c), one can see two sharp characteristic peaks appear at $q=0.3\pi$ and its equivalent position in $1.7\pi$, which confirm a good spiral spin order along the *b* axis as shown in Figs. 2(a) and (b). Furthermore, the non-zero value at $q=0$ indicates the weak ferromagnetism induced by the strong "rung" coupling $J_1$.

After the polarization procedure, $h_e$ is removed and the low temperature calculations of the macroscopic physical properties in a rotating magnetic field *h* are executed, where *h* rotates in a clockwise direction from the *c* axis and in the plane perpendicular to the spiral axis. In LiCu$_2$O$_2$, the anisotropic magnetic phase transition has been observed in the magnetic susceptibility for magnetic field *h*//*ab* and *h*//*c*, respectively. In order to study the anisotropic behavior and compare with the experimental results, we explore the angle dependent of magnetization for *h* rotating in the plane where the spiral spin order lies. Worth to mentioned that it yields the identical results whether *h* rotates in a clockwise or counterclockwise direction. As shown in Fig. 3, all the three components of magnetization follow ***h*** all the time, presenting a well-defined harmonic wave-form. In despite of an exchange anisotropy ($\Delta = 0.7$) in the Hamiltonian (1), the amplitude of $m_c$ curve still exceeds that of $m_a$ and $m_b$. This can be accounted for the strong planar anisotropy in the [110] direction, which ensures the dominated coupling between spins along the *c* axis. Furthermore, $m_a$ and $m_b$ curves show almost no distinction except the sign of the amplitude, being consistent with recent report on untwinned crystals.[15]

Owing to the magnetic origin of ferroelectricity, the ferroelectric behavior in these materials is highly sensitive to the magnetic field. In Fig. 4(a), the angle dependence of the electric polarization $P_c$ under a rotation of ***h*** is displayed. According to the spin current theory, the polarization in this model should emerge along the *c* axis. Our simulation results also confirm this deduction, which is in agreement with the experimental report. [11] As is seen, a hump structure appears in the polarization curve. At *θ*=0 or π where ***h*** is aligned parallel or antiparallel to the *c* axis, $P_c$ presents a narrow valley peak, reaching its minimum. At *θ*=π/2 or 3π/2, namely ***h*** along the vice diagonal of *ab* plane direction, $P_c$ arrives at its maximum, exhibiting a round peak. The varied tendency of $P_c$ in a rating magnetic field can be understood in the following way. From Eq. (3), it is known that the magnitude of the polarization is proportional to $S_iS_j\sin\varphi$ and reaches its maximum at *φ*=π/2, where *φ* denotes the pitch angle between nearest-neighbor spin along the chain. Thus, the results here indicate that the pitch angle increases with ***h*** sweeping from zero to π/2 and decreases with ***h*** sweeping from π/2 to π. When ***h*** sweeps from π to 2π, the behaviors depicted above are repeated again. The duplicated behavior observed in $P_c$ also demonstrates the fact that the exotic magnetic structure of $LiCu_2O_2$ at low temperature is symmetric with respect to *c* and diagonal axes as argued in Ref. [23]. In addition, from the weak $P_c$, it can conjecture that the pitch angle still remains much less than π/2 in the whole process. Moreover, the varied amplitude of $P_c$ is also very small because of a weak magnetic field applied. The angle dependent energy *E* as shown in Fig. 4(b) also verifies the analysis above. It presents the analog picture with that of $P_c$. The minimum of *E* appears at the maximum of pitch angle due to the dominated contribution from the ferromagnetic exchange energy between the nearest-neighbor spins inchain.

In order to obtain a thorough comprehension on the magnetoelectric coupling and microscopic spin structure, the distance $r$ between two nearest-neighbor spin vector is investigated. The sketch map in Fig. 5(a) illustrates the meaning of the physical quantity $r$. It reflects the magnitude of the included angle of two adjacent spins along the chain, which is closely related with the value of the polarization in the system. According to the initial value of $r$ presented in Fig. 5(b), we can obtain a rough estimation on the pitch angle with $h$ applied along the $c$ axis. Its value is at around $\pi/4$. It can also be found that when $h$ is far away from the $c$ axis, the pitch angle increases. However, when $h$ is near the $c$ axis, the variation of the pitch angle is on the contrary. Furthermore, the difference between the maximum and minimum pitch angle is no more than 0.1°. These behaviors indicate the small pitch angle of the spin vector at ground states, and consequently lead to a low-temperature weak electric polarization as shown in Fig. 4(a).

In addition to the anisotropic behaviors, the intriguing field responses are also detected in $LiCu_2O_2$, since it is a complex magnetic order driven multiferroic compound, therefore exploring the influences of various magnetic fields on the magnetoelectric effect will be an effective way to understand the underlying mechanism of magnetism inducing ferroelectricity. Figure 6 presents the variation tendency of the magnetization and polarization under different magnetic fields. As expected, the magnetization is strengthened with the increase of $h$ as shown in Figs. 6 (a) and (b). One can see that the increment of $m_c$ exceeds that of $m_a$ due to the strong spin coupling of z-component, which is also reflected in the $P_c$ curve. In Fig. 6(c), the polarization $P_c$ exhibits large-amplititude variations under a big value of $h$. And compare with the profile in Fig. 4(a), it is greatly smoothened, demonstrating the intimate correlation between the magnetism and ferroelectricity. It reveals that strong magnetic field will be helpful to stabilize the polarization in frustrated system. However, as $h$ increases, $P_c$ is suppressed for $h$ parallel to the $c$ axis or the vice diagonal of $ab$ plane (See $q = 0$ and $\pi/2$), in contrary to the experimental results.[11]

For this discrepancy, the reasons can be attributed to the following several aspects. Based on the calculation of distance *r* in Fig. 5(b), it is known that the pitch angle at ground state is much less than $\pi/2$. When **h**//*c* axis or vice diagonal of *ab* plane, the spin orientation will tend to the direction of the magnetic field. And this tendency will be boost by the enhancement of **h**. As a direct consequence of this variation, the pitch angle decreases, and eventually leads to a decrease of $P_c$ which is proportional to the sine of pitch angle.

In fact, if the ratio of $|J_4/J_2|$ is adjusted to be much greater than 1, the pitch angle larger than $\pi/2$ can be expected at ground states, thereby the variation trend of polarization under different magnetic fields will be similar to that of experimental observation. But it violates the fact that the real value of the pitch angle in $LiCu_2O_2$ is 62.6°.[11] This contradiction manifests the profound effect of quantum fluctuation on the classical incommensurate magnetism, although 3D interactions in real quasi-1D materials tend to suppress the quantum fluctuations. It also implies the discrepancy between our simulation results and those of experiments are mainly ascribe to the neglection of the quantum effects. Therefore, having a deep investigation on the competition between the quantum commensurate and classical incommensurate correlations will be very important and meaningful in understanding the multiferroic nature. In the following work, we will concentrate on this aspect.

## 4. Summary

In summary, the ellipsoidal spiral structure at ground state is produced by performing Monte Carlo simulation. By computing the spin structure factor, the low-temperature spiral order of the system has been confirmed. Focusing on magnetoelectric phenomenon, we have investigated the variations of magnetic and ferroelectric behaviors under rotating **h**. The anisotropic responses of the magnetization and polarization indicate the strong coupling between spins along *c* axis, which is in qualitatively accord with the experimental results. In addition, we also find the strong magnetic field can stabilize the ferroelectricity in frustrated multiferroic compounds, demonstrating the intrinsic connection between magnetism and ferroelectricity in this class of materials. Last but not least, the contrary responses of the polarization under different magnetic fields to the experiments are discussed. The analysis implies the non-neglectable effects of the quantum fluctuations in $LiCu_2O_2$ compound.

**References**

[1] Fiebig M 2005 *J. Phys. D* **38** R123

[2] Spaldin N, Cheong R W and Ramesh R 2010 *Phys. Today* **63** 38

[3] Kimura T, Goto T, Shintani H, Ishizaka K, Arima T and Tokura Y 2003 *Nature (London)* **426** 55

[4] Cheong S W and Mostvoy M 2007 *Nature Mater.* **6** 13

[5] Yamasaki Y, Miyasaka S, Kaneko Y, He J P, Arima T and Tokura Y 2006 *Phys. Rev. Lett.* **96** 207204

[6] Kimura T, Lashley J C and Ramirez A P 2006 *Phys. Rev. B* **73** 220401(R)

[7] Wang K F, Liu J M and Ren Z F 2009 *Adv. Phys.* **58** 321

[8] Enderle M, Mukherjee C, Fäk B, Kremer R K, Broto J M, Rosner H, Drechsler S L, Richter J, Malek J, Prokofiev A, Assmus W, Pujol S, Raggazzoni J L, Rakoto H, Rheinstädter M and Rønnow H M 2005 *Europhys. Lett.* **70** 237

[9] Drechsler S L, Volkova O, Vasiliev A N, Tristan N, Richter J, Schmitt M, Rosner H, Málek J, Klingeler R, Zvyagin A A and Büchner B 2007 *Phys. Rev. Lett.* **98** 077202

[10] Banks M G, Heidrich-Meisner F, Honnecker A, Rakoto H, Broto J M and Kremer R K 2007 *J. Phys. Condens. Matter* **19** 145227

[11] Park S, Choi Y J, Zhang C L and Cheong S W 2007 *Phys. Rev. Lett.* **98** 057601

[12] Bush A A, Glazkov V N, Hagiwara M, Kashiwagi T, Kimura S, Omura K, Prozorova L A, Svistov L E, Vasiliev A M and Zheludev A 2012 *Phys. Rev. B* **85** 054421

[13] Drechsler S L, Málek J, Richter J, Moskvin A S, Gippius A A and Rosner H 2005 *Phys. Rev. Lett.* **94** 039705

[14] Choi K Y, Zvyagin S A, Cao G and Lemmens P 2004 *Phys. Rev. B* **69** 104421

[15] Hsu H C, Liu H L and Chou F C 2008 *Phys. Rev. B* **78** 212401

[16] Masuda T, Zheludev A, Bush A, Markina M and Vasiliev A 2004 *Phys. Rev. Lett.* **92** 177201

[17] Masuda T, Zheludev A, Roessli B, Bush A, Markina M and Vasiliev A 2005 *Phys. Rev. B* **72** 014405

[18] Masuda T, Zheludev A, Roessli B, Bush A, Markina M and Vasiliev A 2005 *Phys. Rev. Lett.* **94** 039706

[19] Mihály L, Dóra B, Ványolos A, Berger H and Forró L 2006 *Phys. Rev. Lett.* **97** 067206

[20] Seki S, Yamasaki Y, Soda M, Matsuura M, Hirota K and Tokura Y 2008 *Phys. Rev. Lett.* **100** 127201

[21] Yasui Y, Sato K, Kobayashi Y and Sato M 2009 *J. Phys. Soc. Jpn.* **78** 084720

[22] Kobayashi Y, Sato K, Yasui Y, Moyoshi T, Sato M and Kakurai K 2009 *J. Phys. Soc. Jpn.* **78** 084721

[23] Zhao L, Yeh K W, Rao S M, Huang T W, Wu P, Chao W H, Ke C T, Wu C E, and Wu M K 2012 *E. P. L.* **97** 37004

[24] Zheng P, Luo J L, Wu D, Su S K, Liu G T, Ma Y C and Chen Z J 2008 *Chin. Phys. Lett.* **9** 3406

[25] Dmitriev D V and Krivnov V Ya 2006 *Phys. Rev. B* **73** 024402

[26] Sirker J 2010 *Phys. Rev. B* **81** 014419

[27] Chen M and Hu C D 2011 *Phys. Rev . B* **84** 094433

[28] Maurice R, Pradipto A M, de Graaf C and Broer R 2012 *Phys. Rev. B* **86** 024411

[29] Krug von Nidda H A, Svistov L E, Eremin M V, Eremina R M, Loidl A, Kataev V, Validov A, Prokofiev A, and Aßmus W 2002 *Phys. Rev. B* **65** 134445

[30] Huang S W, Huang D J, Okamoto J, Mou C Y, Wu W B, Yeh K W, Chen C L, Wu M K, Hsu H C, Chou F C and Chen C T 2008 *Phys. Rev. Lett.* **101** 077205

[31] Qi Y and Du A 2013 arXiv:1312.0293 [cond-mat.mtrl-sci]

[32] Yao X 2011 *E. P. L.* **94** 67003

**Figure Captions**

Fig. 1: (Color online) A schematic view of exchange interactions between magnetic $Cu^{2+}$ ions in $LiCu_2O_2$.

Fig. 2: (a) The snapshot for the coordinates of spins in all zigzag chains after the system polarized by a small electric field. (b) The spin configurations for partial spins in a single chain. (c) The wave vector dependence of the spin structure factor .

Fig. 3: (Color online) The angle dependent magnetization under $|h|$=1.

Fig. 4: The angle dependent (a) polarization $P_c$ and (b) energy $E$ under $|h|$=1.

Fig. 5: (a) The sketch map of the spin structure in ground state, where the distance between the nearest-neighbor spin vectors along the chain is denoted as $r$. (b) The angle dependent $r$ under $|h|$=1.

Fig. 6: (Color online) The angle dependent (a) magnetization $m_a$, (b) magnetization $m_c$ and (c) polarization $P_c$ under different magnitudes of $h$.

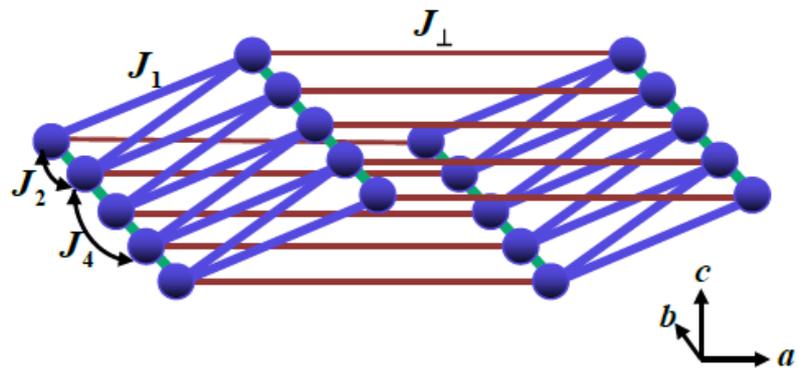

Figure 1

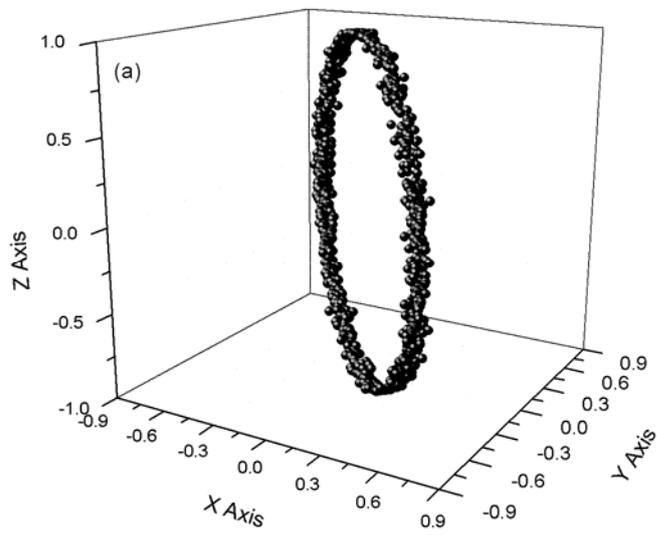

Figure 2 (a)

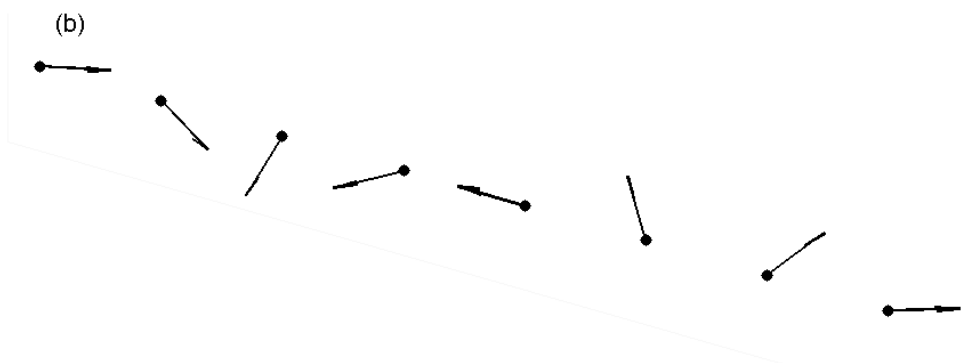

Figure 2 (b)

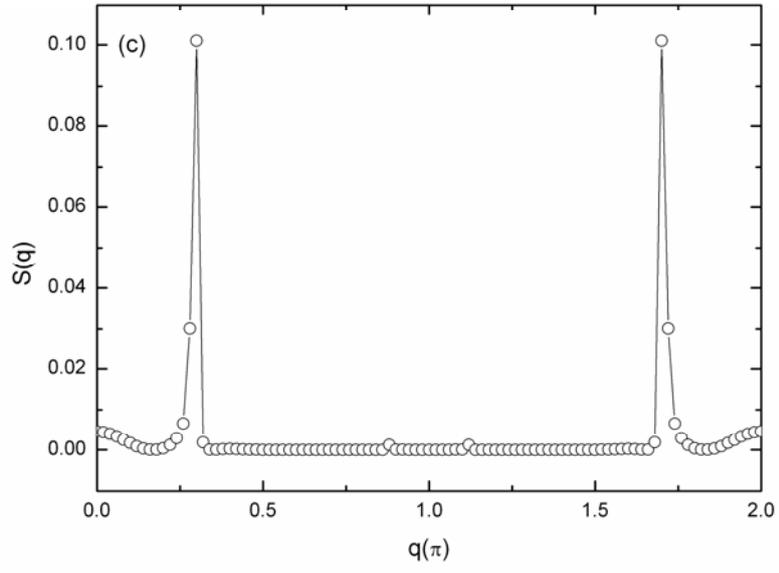

Figure 2 (c)

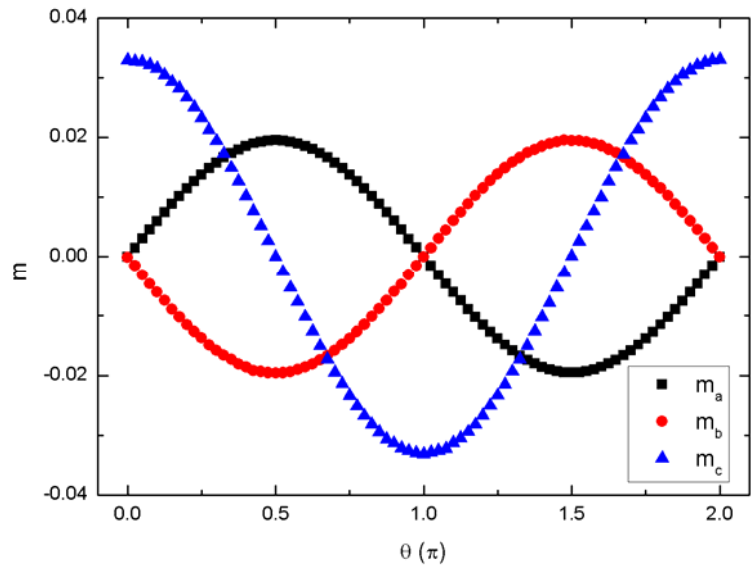

Figure 3

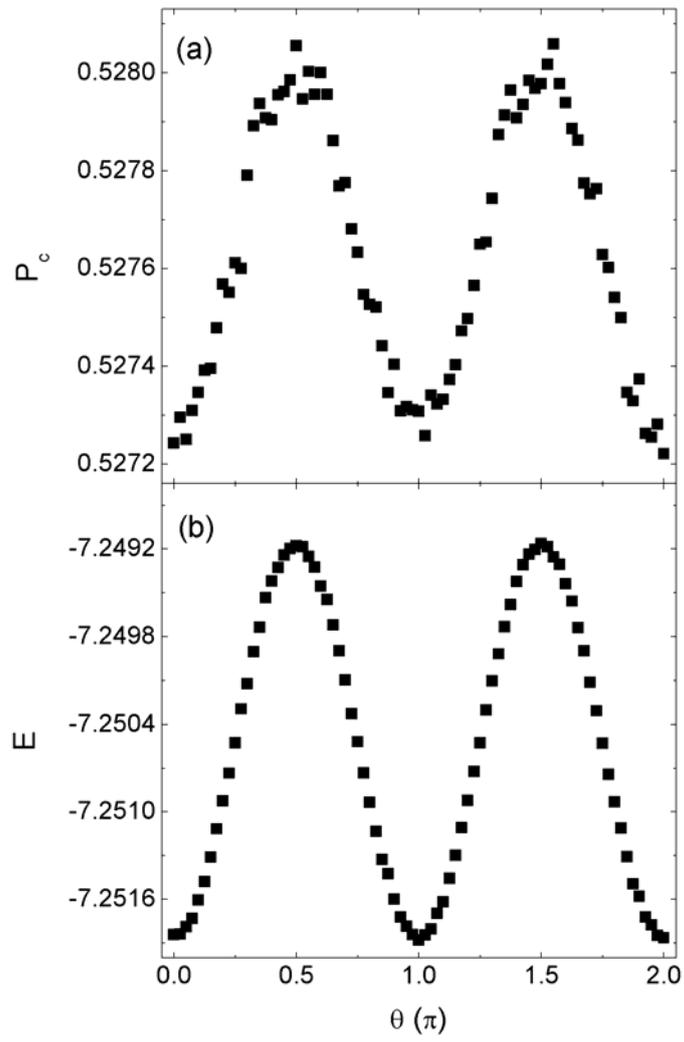

Figure 4

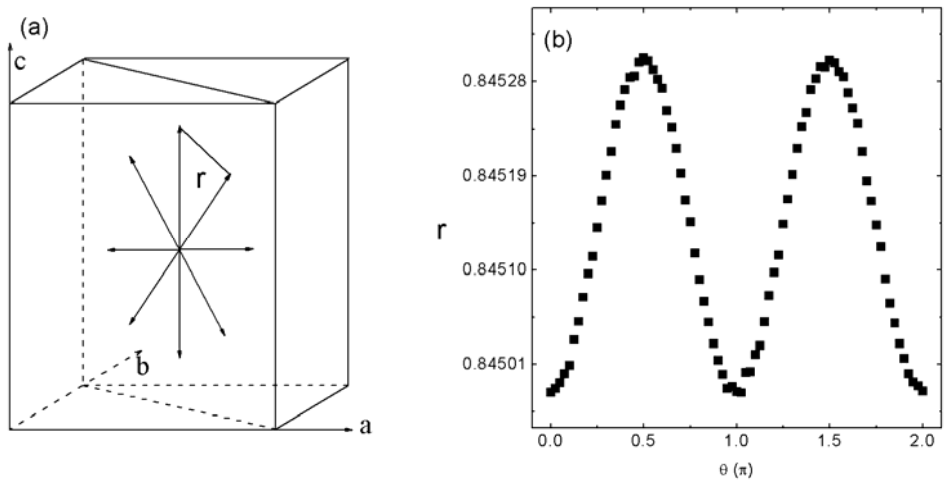

Figure 5

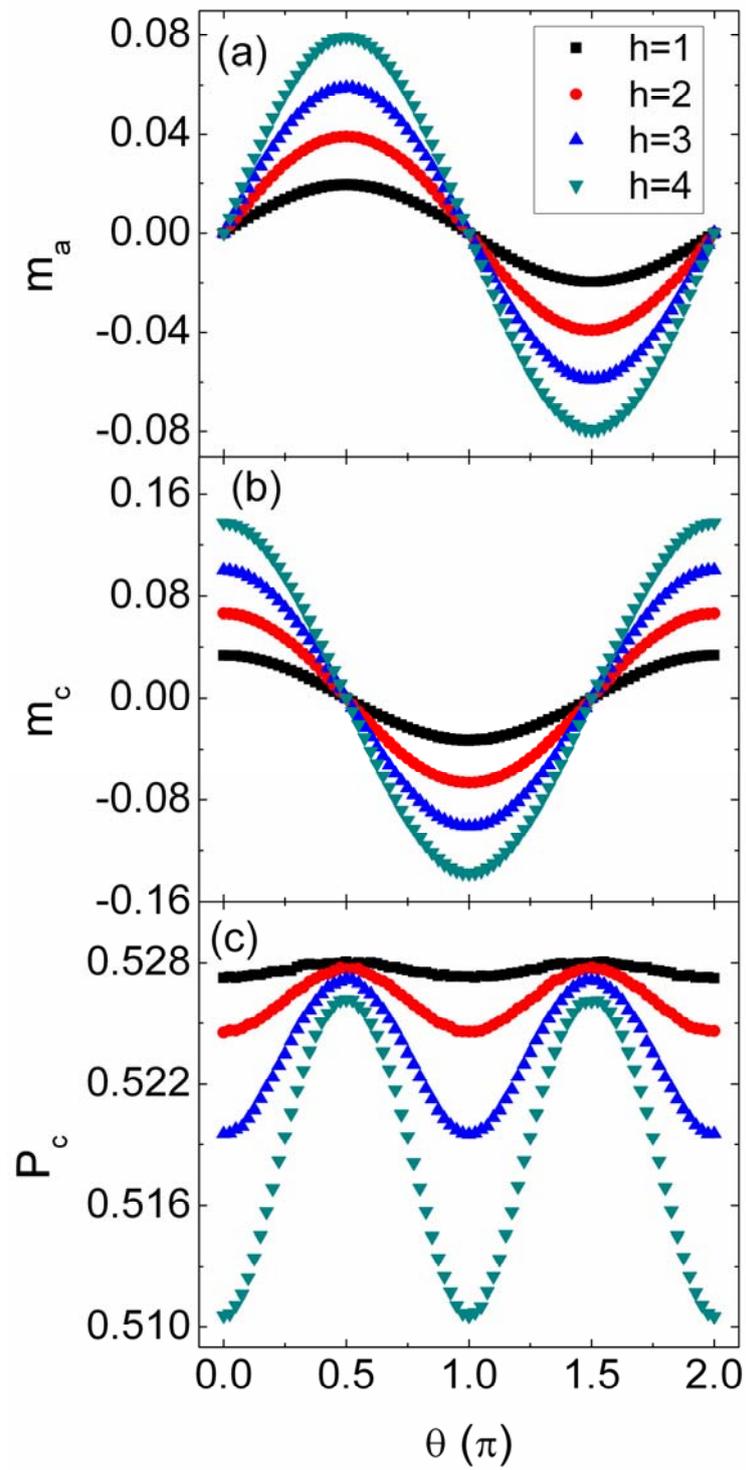

Figure 6